\documentclass[showpacs,preprintnumbers,prd,nofootinbib,floats,amssymb,floatfix]{revtex4}
\usepackage{amsmath}
\usepackage{graphicx}
\usepackage{amsxtra}
\usepackage{hyperref}
\usepackage{amsmath}
\usepackage{amstext}
\begin{document}
\title{ Hamilton-Jacobi analysis for  three dimensional gravity without dynamics}

\author{Alberto Escalante}  \email{aescalan@ifuap.buap.mx}
\author{ I.  Vallejo-Fabila}  \email{ivallejo@ifuap.buap.mx}
 \affiliation{  Instituto de F{\'i}sica, Benem\'erita Universidad Aut\'onoma de Puebla. \\
 Apartado Postal J-48 72570, Puebla Pue., M\'exico, }
\begin{abstract}
The Hamilton-Jacobi analysis for gravity without dynamics is performed. We report a detailed analysis where the complete set of Hamilton-Jacobi constraints,  the characteristic equations and the gauge transformations of the theory are found. We compare our results with those reported in the literature where alternative approaches are used. In addition,  we complete our work  by performing the canonical covariant analysis by constructing    a gauge invariant symplectic structure,   and  we find a full consistency between the results obtained from both approaches.
\end{abstract}
 \date{\today}
\pacs{98.80.-k,98.80.Cq}
\preprint{}
\maketitle
\section{Introduction}
It is well known that three dimensional gravity [3dg] described by Palatini's first order formulation  can be expressed as a full connection theory, this is, under a correct election of a connection in a Chern-Simons [CS] theory,   3dg  can be obtained. In fact, it has been showed that these theories are equivalent at  Lagrangian level up to a total derivative \cite{1, 2}. Furthermore,  that CS theory turns out to be  an extension of three dimensional gravity,   since  the later requires an invertible triad,  whereas in the former  this is not a necessary ingredient  \cite{3}. Hence,  the study of CS  theories has been a topic of great interest for the theoretical physics community,  because  these theories  could  help us to understand  the classical and quantum connection between real gravity and gauge theories \cite{1, 2, 3, 4}.  In this respect, the analysis of toy models such as CS theories  are great laboratories  for  testing ideas that could be applicable in real gravity  in either classical and   quantum regime. Furthermore, the relation between CS  and 3dg  theories can be extended in order to obtain models  with  more general structure than Palatini's theory, for instance  the Bonzom Livine  model [BL] \cite{5, 6}   and the models  reported by  V. Hussain,    in which  there is not generator of the dynamics \cite{7, 8}. The BL model  describes a set of actions sharing the equations of motion with Palatini's theory; however, the symplectic structure in the BL model depends on a Barbero-Immirizi-like parameter, which may represent a difference at dynamical level \cite{5, 6}.  In contrast  to   real gravity theory, on the other hand,  in the  Hussain  theories  have  not a Hamiltonian constraint,  but the vector  and the Gauss constraints are present   and this fact   facilitates    the  study of   quantum aspects of gravity being it   a difficult task to perform, thus, the study of toy models  brings  us insights  for study the symmetries of gravity. In this respect, there is an  alternative way  for  studying  the symmetries of gravity theories, it  is based on the Dirac formulation which turns on to be the cornerstone of the modern formulation of canonical gravity,  the so-called Loop Quantum Gravity  \cite{9, 10, 11, 12, 13, 14}.  In general, the Dirac method \cite{15, 16}  is an alternative  for studying the symmetries of singular systems; it brings  us  the identification of the constraints of the theory; which  can be of first class (gauge generators),  or second class (blocks for constructing the Dirac brackets), and then one can identify  relevant information, such as   the counting the physical degrees of freedom, the gauge symmetries   and  other  symmetries.  In spite of the Dirac formulation is an elegant way for studying gauge systems,  the development  of  this framework step by step generally  is a large and tedious task, hence, it is necessary to use alternative formulations that could give us a complete canonical description of the theory;  in this respect,   there exist  also alternative methods for studying singular systems,  the so-called Faddeev-Jackiw [FJ], the canonical covariant and the Hamilton-Jacobi [HJ] formulations. The FJ  framework is a symplectic approach \cite{17, 18, 19}; all relevant symmetries of the theory  can be obtained through a symplectic tensor and the  symplectic variables identified as the degrees of freedom. Furthermore, in FJ method all constraints of the theory are to be treated on the same footing, say, it is not necessary to perform the classification of the constraints in  first class or second class such as in Dirac'€™s method is done. In addition, if the  symplectic tensor  is obtained, then its components are identified with the FJ generalized brackets; at the end Dirac'€™s brackets and FJ brackets coincide. The canonical covariant method \cite{20}, on the other hand, is based on the construction of a closed and gauge invariant geometric structure defined  on the covariant phase space. From the symplectic structure,  we are able to identify the symmetries of the theory such as canonical  and  gauge transformations. Finally, the $HJ$ formalism is an interesting framework for studying singular systems, it was  developed by  G\"uller and   based on the Carath\'eodory ideas \cite{21, 22}. In general, in the $HJ$ framework  a fundamental differential  is constructed, from which it is possible to identify the complete set of constraints  called Hamiltonians, however, the   $HJ$ formalism has  advantages  in the  identification of the Hamiltonians,  because no analogue of the Dirac conjecture is needed. Once we have identified all the Hamiltonians, it is possible to identify the gauge transformations,  the generalized $HJ$  brackets  (which will coincide with the Dirac brackets)  and the characteristic equations that describe the dynamical evolution of the system.  Thus, the $HJ$ framework is complete by itself  with advantages  respect to other formulations \cite{23, 24}.\\
In this manner, with the antecedents mentioned above, in this paper we will apply the $HJ$ method   to 3dg  without dynamics. The model proposed by  Hussain has been studied in \cite{7, 8, 8a} by using the Dirac  and  the FJ formulations. However, we are interested in to perform  a detailed $HJ$ formalism in order to report the symmetries of the theory under study from a different point of view to those reported in \cite{8a}. On the other hand,  a complete $HJ$ analysis can be an interesting element with insights to  be extended to other models with a more complex structure  than  Chern-Simons gravity such as 3d massive gravity,   or topological theories with boundaries \cite{25}.  In this respect, the $HJ$ method could be a sophisticated alternative for studying gauge systems with general covariance. Finally, we have added the canonical  covariant analysis, in order to complete the study by using all schemes reported in the literature. In particular, we shall construct a gauge invariant symplectic structure from which we  reproduce the results obtained from  the $HJ$ method.\\
The paper is structured as follows. In the Section II   a  detailed $HJ$ analysis for  3dg without dynamics is performed. From the fundamental differential, we identify the complete set of Hamiltonians, then we report the characteristic equations  and the gauge transformations. In the Section III  the canonical covariant method is developed; from a closed and gauge invariant two form, we reproduce  results  obtained in previous sections. We finish in the Section IV with some remarks and conclusions
\section{Hamilton-Jacobi analysis}
As  it was   commented  previously, a  gravity theory can be obtained from a CS theory. In fact, if we choice to work with the following generators $P_{i}$ and $J$ satisfying the commutation rules,
\begin{align}
\left[P_{i}, P_{j}\right] = \Lambda\epsilon_{ij}J, &\quad 	\left[J_{i}, P_{i}\right] = \epsilon_{ij}P^{j},
\end{align}
where the internal indices $i,j,k=1,2$ are raised by the $2D$ metric $(+,+)$ for $\Lambda$ positive and a the internal group   SO(3),  and $(-,+)$ for $\Lambda$ negative and  with an  internal group  SO(2,1);  in this paper we will work with a positive cosmological constant $\Lambda$. Hence, by considering the connection $A_{\mu} = e^{i}{_{\mu}}P_{i} + \omega_{\mu}J$    in the following Chern-Simons action
\begin{equation}
S[A] = \int Tr\left(A\wedge dA + \frac{2}{3}A\wedge A\wedge A\right),
\end{equation}
where $\mu,\alpha,\beta =0,1,2$, with the traces  $<J,J>=1, <P_{i}, P_{j}>= \Lambda J_{ij}$,  $<J, P_{i}>=0, $ and $ <A\wedge A\wedge A>=\frac{1}{2}<A, [A, A]>$, then the  following action arises 
\begin{equation}
S\left[e,\omega\right]= \int \epsilon^{\alpha\beta\gamma}\left[\omega_{\alpha\partial_{\beta}\omega_{\gamma}} + \Lambda e^{i}{_{\alpha}}\partial_{\beta}e_{\gamma i} + \Lambda\epsilon_{ij}e^{i}{_{\alpha}}e^{j}{_{\beta}}\omega_{\gamma}\right]dx^{3},
\label{acti}
\end{equation}
this is the action reported in \cite{7}  and analyzed in \cite{7, 8a} by using the FJ  and Dirac analysis. In order to extend  these works,  we will  perform the  $HJ$ and the canonical covariant methods. \\
We start by performing the $2+1$ decomposition,
\begin{align}
S\left[e,\omega\right] &= \int \left[\Lambda\epsilon^{ab}e^{j}{_{b}}\dot{e}_{ai} + \epsilon^{ab}\omega_{b}\dot{\omega}_{a} + \omega_{0}\{2\epsilon^{ab}\partial_{b}\omega_{b} + \Lambda\epsilon^{ab}\epsilon_{ij}e^{i}{_{a}}e^{j}{_{b}}\}\right] \\
&+\left[ e^{i}{_{0}}\{2\Lambda\epsilon^{ab}\partial_{a}e_{bi} + 2\Lambda\epsilon^{ab}\epsilon_{ij}e^{j}{_{a}}\omega_{b}\}\right]dx^{3},
\end{align}
where we defined  $\epsilon^{0ab}\equiv\epsilon^{ab}$. Furthermore, we identify the canonical momenta $\Pi^{\mu}, P^{\mu}{_{i}}$ conjugated to $\omega_{\mu}, e^{i}{_{\mu}}$ 
\begin{align*}
\Pi^{\mu} &\equiv \frac{\partial\mathcal{L}}{\partial\dot{\omega}_{\mu}}, & 	
P^{\mu}{_{i}} \equiv  \frac{\partial\mathcal{L}}{\partial\dot{e}^{i}{_{\mu}}}, 
\end{align*}
thus, we identify the following Hamiltonians
\begin{align}
H'&\equiv\Pi + H_{0} = 0, \nonumber \\
\bar{g}^{a}&\equiv\Pi^{a} - \epsilon^{ab}\omega_{b} = 0, \nonumber \\
\bar{g}^{0}&\equiv\Pi^{0} = 0, \nonumber\\
\bar{D}^{0}{_{i}}&\equiv P^{0}{_{i}} = 0, \nonumber\\
\bar{D}^{a}{_{j}}&\equiv P^{a}{_{j}} - \Lambda\epsilon^{ab}e_{bj} = 0,
\label{6}
\end{align}
and the following fundamental Poisson brackets are identified from the definition of the momenta
\begin{align*}
\{e^{i}{_{\mu}}, P^{\alpha}{_{j}}\} &= \delta^{\alpha}{_{\mu}}\delta^{i}{_{j}}\delta^{2}(x-y), \\
\{\omega_{\mu}, \Pi^{\alpha}\} &= \delta^{\alpha}{_{\mu}}\delta^{2}(x-y),
\end{align*}
where we have defined $\Pi\equiv\partial_{0}S$, with $S$ being the action and $H_{0}$ is the canonical Hamiltonian given by
\begin{align}
H_{0} &= -\omega_{0}\{2\epsilon^{ab}\partial_{a}\omega_{b} + \Lambda\epsilon^{ab}\epsilon_{ij}e^{i}{_{a}}e^{j}{_{b}}\} \\
&- e^{i}{_{0}}\{2\Lambda\epsilon^{ab}\partial_{a}e_{bi} + 2\Lambda\epsilon^{ab}\epsilon_{ij}e^{j}{_{a}}\omega_{b}\},
\end{align}

in terms of the canonical momenta $H_0$ can be written as 
\begin{align}
H_{0} &= -\omega_{0}\{2\partial_{a}P^{a} + \epsilon_{ij}e^{i}{_{a}}\Pi^{aj}\} - e^{i}{_{0}}\{2\partial_{a}\Pi^{a}{_{i}} + 2\Lambda\epsilon_{ij}e^{j}{_{a}}P^{a}\}.
\end{align}
Now with the $HJ$ Hamiltonians we construct the following fundamental $HJ$ differential \cite{21, 22, 23, 24}
\begin{align}
df(x) &=\int d^{3}y\left(\{f(x), H'\}dt + \{f(x), \bar{g}^{\mu}\}d\xi_{\mu} + \{f(x),\bar{D}^{\mu}{_{i}}\}d\zeta^{i}{_{\mu}}\right),
\end{align}
where $\xi_\mu$ and $\zeta^{i}{_{\mu}}$ are parameters related to the $HJ$ constraints $\bar{g}^{\mu}$ and $\bar{D}^{\mu}{_{i}} $ respectively.  Furthermore, all Hamiltonians having vanishing Poisson brackets to each other   are called as  involutives; in the Dirac terminology they are first class constraints. Otherwise, they are non-involutive Hamiltonians (second class constraints). We can see that $\Pi^{0}$ and $P^{i}{_{0}}$  are  involutive Hamiltonians  and $\bar{g}^{a}$ and $\bar{D}^{a}{_{i}}$  are  non-involutives. In this manner, the matrix whose entries are the Poisson brackets between the non-involutive Hamiltonians  is given by
\begin{eqnarray*}
\label{eq}
C_{\alpha\beta}=
\left(
  \begin{array}{cccc}
  -2\Lambda\epsilon^{ag}\eta_{ij}	&\quad		0\\
0	&\quad		-2\epsilon^{ab}	\\
 \end{array}
\right) \delta^{2}(x-y),
\end{eqnarray*}

and its inverse reads 

\begin{eqnarray*}
\label{eq}
(C{_{\alpha\beta}})^{-1}=
\left(
  \begin{array}{cccc}
  \frac{1}{2\Lambda}\epsilon_{ag}\eta^{ij}	&\quad		0\\
0	&\quad		\frac{1}{2}\epsilon_{ab}	\\
 \end{array}
\right) \delta^{2}(x-y),
\end{eqnarray*}
thus, the inverse of the $C$ matrix  allows  us to introduce  generalized brackets  given by \cite{21, 22, 23, 24}
\begin{align}
\{A, B\}^{*} &=\{A, B\} - \{A, H'_{\bar{a}}\}(C{_{\bar{a}\bar{b}}})^{-1}\{H'_{\bar{b}}, B\},
\label{10}
\end{align}
where $H'_{\bar{a}}$ are the non-involutive Hamiltonians. In this manner, by using (\ref{10}) the non-zero generalized brackets are given by
\begin{eqnarray}
\{e^{i}{_{a}}(x), e^{j}{_{b}}(y)\}^{*} &=& \frac{1}{2\Lambda}\epsilon_{ab}\eta^{ij}\delta^{2}(x-y), \nonumber \\
\{\omega_{a}(x), \omega_{b}(y)\}^{*} &=& \frac{\epsilon_{ab}}{2}\delta^{2}(x-y),  \nonumber \\
\{e^{i}{_{a}}(x), \Pi^{b}{_{j}}(y)\}^{*} &=& \frac{1}{2}\delta^{a}{_{b}}\delta^{i}{_{j}}\delta^{2}(x-y),	\nonumber \\
\{\Pi^{a}{_{i}}(x), \Pi^{b}{_{j}}(y)\}^{*} &=& \frac{\epsilon^{ab}}{2}\eta_{ij}\delta^{2}(x-y),	 \nonumber \\
\{P^{a}(x), P^{b}(y)\}^{*} &=& \frac{\epsilon^{ab}}{2}\delta^{2}(x-y),	\nonumber \\
\{\omega_{a}(x), P^{b}(y)\}^{*} &=& \frac{\delta^{a}{_{b}}}{2}\delta^{2}(x-y). 
\label{bra}
\end{eqnarray}
The introduction of the generalized brackets redefine the dynamics. In fact,  the non-involutive constraints are removed  from the fundamental differential and  it can be   expressed  in terms of the generalized brackets and involutive Hamiltonians, 
\begin{align}
df(x) &= \int d^{2}y\left(\{f(x), H'(y)\}^{*}dt + \{f(x), \Pi^{0}(y)\}^{*}d\xi_{0} + \{f(x), P^{0}{_{i}}(y)\}^{*}d\zeta^{i}{_{0}}\right),
\end{align}
thus,  the \textit{Frobenius integrability} conditions for the involutive Hamiltonians \cite{21, 22, 23, 24}, say  $\Pi^{0}$  and  $P^{0}{_{i}}$, introduce new Hamiltonians 
\begin{align*}
d\Pi^{0} &= \int d^{2}y\left(\{\Pi^{0}(x), H'(y)\}^{*}dt + \{\Pi^{0}(x), \Pi^{0}(y)\}^{*}d\xi_{0} + \{\Pi^{0}(x), P^{0}{_{i}}(y)\}^{*}d\zeta^{j}{_{0}}\right) = 0 \\
&= \partial_{a}P^{a} + \frac{1}{2}\epsilon_{ij}e^{i}{_{a}}\Pi^{aj} =0 \\
dP^{0}{_{i}} &= \int d^{2}y\left(\{Pi^{0}(x), H'(y)\}^{*}dt + \{P^{0}{_{i}}(x), \Pi^{0}(y)\}^{*}d\xi_{0} + \{P^{0}{_{i}}(x), P^{0}{_{j}}(y)\}^{*}d\zeta^{j}{_{0}}\right) = 0	\\
&= \partial_{a}\Pi^{a}{_{i}} + \Lambda\epsilon_{ij}e^{j}{_{a}}P^{aj} =0.
\end{align*}
Hence, by using (\ref{6})   the new Hamiltonians can be written as
\begin{align}
C&\equiv\epsilon^{ab}\partial_{a}\omega_{b} + \frac{\Lambda}{2}\epsilon_{ij}e^{i}{_{a}}e^{j}{_{b}},	\nonumber \\
C_{i}&\equiv\epsilon^{ab}\partial_{a}e_{bi} + \epsilon^{ab}\epsilon_{ij}e^{j}{_{a}}\omega_{b}.
\label{cons}
\end{align}
Moreover, the generalized algebra of these new Hamiltonians is given by 
\begin{align*}
\{C(x), C(y)\}^{*} &=0,		\\
\{C(x), C^{i}(y)\}^{*} &= \frac{\epsilon^{i}{_{j}}}{2}C^{j},		\\
\{C^{i}(x), C^{j}(y)\}^{*} &= \frac{\epsilon^{ij}}{2\Lambda}C,
\end{align*}
where we observe that $C$ and $C^i$ are involutive Hamiltonians. Moreover, the Hamiltonian $C$ is the equivalent one to the Gauss constraint and it generates  Abelian transformations on the $\omega_a$ and rotations on the $e$ field, whereas the Hamiltonian $C_i$ is the equivalent to the vector constraint.  The   algebra is closed and  this allows us conclude that  there are not more  new Hamiltonians.  In this manner,   we construct the following fundamental differential in terms of  all involutive Hamiltonians, 
\begin{eqnarray}
df(x) &= &\int d^{2}y [ \{f(x), H'(y)\}^{*}dt + \{f(x), \Pi^{0}(y)\}^{*}d\xi_{0} \nonumber \\
&+ & \{ f(x), P^{0}{_{i}}(y)\}^{*}d\zeta^{i}{_{0}} + \{f(x), C(y)\}^{*}d\lambda + \{f(x), C^{i}(y)\}d\lambda_{i}] ,
\label{13}
\end{eqnarray}
here, $\lambda$ and $\lambda_i$ are parameters related to the Hamiltonians $C$ and $C^i$  respectively. From the fundamental differential (\ref{13}) we can calculate the characteristic equations,  this is,  those equations  governing  the evolution of the dynamical variables  
\begin{align}
d\omega_{0} &= d\xi_{0}, \nonumber 	\\
de^{i}{_{0}} &= d\zeta^{i}{_{0}},	 \nonumber  \\
d\omega_{a} &= \Big(\partial_{a}\omega_{0} - e^i_0 \Lambda \epsilon_{ij} e^j _a \Big)dt - \frac{\partial_{a}}{2}d\lambda + \epsilon^{k}{_{l}}\frac{e^{l}{_{a}}}{2}d\lambda_{k}, \nonumber \\
d e^{i}{_{a}}&=\left(\partial_a e^i_0+  \omega_0 \epsilon^i{_{j}} e^j_a + e^k_0\epsilon_{k} {^{{i}}} \omega_a  \right)dt + \epsilon^{i}{_{k}}e^{k}{_{a}}d\lambda - \frac{\partial_{a}}{2\Lambda}d\lambda^{i} - \frac{1}{2\Lambda}\omega_{a}\epsilon^{ki}d\lambda_{k}.
\end{align}
from these equations we observe that the fields $\omega_0$ and $e^i_0$ are Lagrange multipliers. On the other hand, from the characteristic equations we can identify  the equations of motion of the theory given by 
\begin{eqnarray}
\partial_0 \omega_{a} &=&\partial_{a}\omega_{0} - e^i_0 \Lambda \epsilon_{ij} e^j _a,  \nonumber \\
\partial_0 e^{i}{_{a}}&=& \partial_a e^i_0+  \omega_0 \epsilon^i{_{j}} e^j_a + e^k_0\epsilon_{k} {^{{i}}} \omega_a. 
\end{eqnarray}
Furthermore, we also identify  important symmetries, for instance, by considering $dt=0$ in the characteristic equations, we obtain
\begin{eqnarray}
\delta\omega_{0} &=& \delta\xi_{0},  \nonumber \\
\delta e^{i}{_{0}} &=& \delta\zeta^{i}{_{0}}, \nonumber \\
\delta\omega_{a} &= &\partial_{a} \delta\lambda + \epsilon{_{l}}^k{e^{l}_{{a}}}\delta\lambda_{k}, \nonumber \\
\delta e^{i}{_{a}} &= &\partial_{a}\delta\lambda^{i}- \omega_{a}\epsilon^{ik}\delta \lambda_{k} + \Lambda \epsilon^{i}{_{k}}e^{k}{_{a}}\delta\lambda ,  
\label{trans2}
\end{eqnarray}
that correspond to the gauge transformations reported in \cite{8a}. We can observe that  all results reported in \cite{8a} have been reproduced  in more economical way in the $HJ$  approach, thus, the $HJ$ formalism is more direct  than  both Dirac's  and FJ approaches.  In this respect,  the $HJ$ method is a good alternative for studying gauge systems. \\
We finish this section by carrying out the counting of physical degrees of freedom. We have seen from the fundamental differential all dynamics of the system is given in terms of the dynamical variables, the involutive Hamiltonians   and the remaining number of parameters associated to the Hamiltonians. In this case,  the dynamical variables can be identified from the characteristic equations and they are given by 2 $\omega_a$ and 6 $e^i_a$ fields;  the  relevant Hamiltonians are the  four  involutive ones $C$ and $C^i$, and  there are four parameters  related to the involutive Hamiltonians  $\lambda$ and $\lambda_k$. In this manner, the degrees of freedom are $DF= 8- 4-4=0$, thus the theory is devoid of physical degrees of freedom,  as expected.    
\section{The symplectic method for gravity without dynamics}
Now, in order to complete our analysis,   we will perform the canonical covariant formalism.  Our starting point is again the action (\ref{acti}), and the variation of the action with respect the dynamical variables  $\omega_{\alpha}$ and $e^{i}{_{\alpha}}$  reads 
\begin{align}
\delta [S]  &=\int_{\mu}\epsilon^{\alpha\beta\gamma}\Big[\left(2\partial_{\beta}\omega_{\gamma} - \Lambda\epsilon_{ij}e^{i}{_{\gamma}}e^{i}{_{\beta}}\right)\delta\omega_{\alpha} + \left(2\Lambda\partial_{\beta}e_{i\gamma} + 2\Lambda\epsilon_{ij}e^{j}{_{\beta}}\omega_{\gamma}\right)\delta e^{i}{_{\alpha}} \nonumber  \\
&+ \partial_{\beta}\left(\omega_{\alpha}\delta\omega_{\gamma} + \Lambda e^{i}{_{\alpha}}\delta e_{\gamma}\right)\Big]dx^{3},
\label{lin1}
\end{align}
where we can identify the equations of motion
\begin{align}
\epsilon^{\alpha\beta\gamma}\left(\partial_{\beta}\omega_{\gamma} - \frac{\Lambda}{2}\epsilon_{ij}e^{i}{_{\gamma}}e^{j}{_{\beta}}\right) &= 0, 
\label{mot1}
\end{align}
\begin{align}
\epsilon^{\alpha\beta\gamma}\left(\partial_{\beta}e_{i\gamma} + \epsilon_{ij}e^{j}{_{\beta}}\omega_{\gamma}\right) &= 0, 
\label{mot2}
\end{align}
moreover, from the total divergence present in (\ref{lin1}) we identify the symplectic potential of the theory \cite{21a}
\begin{align*}
\psi^{\beta} &= \epsilon^{\beta\alpha\gamma}\left(\omega_{\alpha}\delta\omega_{\gamma} + \Lambda e^{i}{_{\alpha}}\delta e _{\alpha i}\right).
\end{align*}
No we define the cornerstone of the canonical covariant method,   the covariant phase space; the covariant phase space for the theory described by (\ref{acti}) is the space of solutions of (\ref{mot1}), (\ref{mot2}), and we will call it Z \cite{20}. Hence, on Z the fields $\omega_a$ and $e^i_a$ are zero-forms and its variations $ \delta\omega_{\alpha}$ and $ \delta e^{i}{_{\alpha}}$ are 1-forms. On the other hand, the linearized equations of motion,  useful for future calculations,  are given by replacing in the equations of motion,  
$\omega_{\alpha}\longrightarrow  \omega_{\alpha} + \delta\omega_{\alpha}$,  and $e^{i}{_{\alpha}}\longrightarrow e^{i}{_{\alpha}} + \delta e^{i}{_{\alpha}}$,  and keeping only the first order terms, we obtain
\begin{align}
\epsilon^{\alpha\beta\gamma}\left(\partial_{\beta}\delta\omega_{\gamma} - \Lambda\epsilon_{ij}e^{i}{_{\gamma}}\delta e^{j}{_{\beta}}\right) &= 0,  \nonumber \\
\epsilon^{\alpha\beta\gamma}\left(\partial_{\beta}\delta e_{\gamma i} + \epsilon_{ij}e^{j}{_{\beta}}\delta\omega_{\gamma} + \epsilon_{ij}\delta e^{j}{_{\beta}}\omega_{\gamma}\right) &= 0.
\label{lin}
\end{align}
In this manner the exterior derivative of the symplectic potential gives the symplectic structure of the theory, this is 
\begin{align}
\bar{\omega} &= \int J^{\mu}d\Sigma_{\mu} = \int\delta\psi^{\mu}d\Sigma_{\mu} = \int_{\Sigma}\epsilon^{\beta\gamma\alpha}\left(\delta\omega_{\gamma}\wedge\delta\omega_{\alpha} + \Lambda\delta e^{i}{_{\gamma}}\wedge\delta e_{i\alpha}\right)d\Sigma_{\beta},
\end{align}
Where $\Sigma$ is a Cauchy surface. We can observe in the symplectic structure the noncommutative character of the fields $\omega$ and $e$  (see the generalized brackets (\ref{bra})).\\
On the other hand, we will prove that $\bar{\omega}$ is closed and gauge invariant; the closeness of $\bar{\omega}$ is equivalent one to the Jacobi identity that Poisson brackets satisfy in the  Hamiltonian scheme. On the other hand, the gauge invariance is reflection  of the symmetries of the theory.     Furthermore, the integral kernel of the geometric form,  say $J^{\mu}$, is conserved. This fact is important because it guarantees that $\bar{\omega}$ is independient of $\Sigma$. Hence, we observe that $\delta^{2}e^{i}{_{\mu}} = 0$ and $\delta^{2}\omega_{\alpha}=0$, because $e^{i}{_{\mu}}$ and $\omega_{\alpha}$ are independent zero forms on $Z$ and $\delta$ is nilpotent, thus
\begin{align}
\delta\bar{\omega} &= \int_{Z}\epsilon^{\beta\gamma\alpha}\left(\delta^{2}\omega_{\gamma}\wedge\delta\omega_{\alpha} - \delta\omega_{\gamma}\wedge\delta^{2}\omega_{\alpha} + \Lambda\delta^{2}e^{i}{_{\gamma}}\wedge\delta e_{i\alpha} - \Lambda e^{i}{_{\gamma}}\wedge\delta^{2}e_{i\alpha}\right)d\Sigma_{\beta} =0,
\end{align}
this indicates  that $\bar{\omega}$ is closed. Now, we can observe that under fundamental gauge transformations generating rotations given in (\ref{trans}),  and for some infinitesimal variation we have
\begin{eqnarray}
\delta e'^{i}{_{\alpha}} &= &\delta e^{i}{_{\alpha}} + \Lambda \epsilon^{i}{_{k}}\epsilon\delta e^{k}{_{\alpha}}, \nonumber \\
\delta\omega'_{\alpha} &=&\delta\omega_{\alpha},
\label{24a}
\end{eqnarray}
thus, by using (\ref{24a}) we observe that $\bar{\omega}$ transforms
\begin{align}
\bar{\omega'} &= \int_{\Sigma}\epsilon^{\beta\alpha\gamma}\left(\delta\omega'_{\alpha}\wedge\delta\omega'_{\gamma} + \Lambda\delta e'^{i}{_{\alpha}}\wedge\delta e'_{i\gamma}\right)d\Sigma_{\beta} =
\bar{\omega} + \int_{\Sigma}O(0^{2})d\Sigma,
\end{align}
therefore, $\bar{\omega}$ is a $SO(3)$ singlet. This fact allows us to prove that 
\begin{align}
\partial_{\mu}J^{\mu} &= \epsilon^{\mu\alpha\gamma}\left(\partial_{\mu}\delta\omega_{\alpha}\wedge\delta\omega_{\gamma} + \delta\omega_{\alpha}\wedge\partial_{\mu}\delta\omega_{\gamma} + \Lambda\partial_{\mu}\delta e^{i}{_{\alpha}}\wedge\delta e{_{i\gamma}} + \Lambda\delta e^{i}{_{\alpha}}\wedge\partial_{\mu}\delta e_{i\gamma}\right) 	\nonumber	\\
&= -2\epsilon^{\mu\alpha\gamma}\Lambda\epsilon_{ij}e^{i}{_{\mu}}\delta e^{j}{_{\alpha}}\wedge\delta\omega_{\gamma} - 2\epsilon^{\mu\alpha\gamma}\epsilon_{ij}e^{j}{_{\mu}}\delta\omega_{\alpha}\wedge\delta e_{i\gamma} \nonumber	\\
& -2\Lambda\epsilon^{\mu\alpha\gamma}\epsilon_{ij}\omega_{\alpha}\delta e^{j}{_{\mu}}\wedge\delta e_{i\gamma} \nonumber	\\
&= 0,
\end{align}
where we have used the antisymetry of 1-forms $\delta e^{i}{_{\alpha}}$ and $\delta\omega_{\alpha}$, and the linearized equations of motion (\ref{lin}). Therefore $\bar{\omega}$ is independient of $\Sigma$.\\
On the other hand, we know that the theory is diffeomophisms covariant, in this manner, for some infinitesimal variation the gauge transformations (\ref{trans2}) take the form
\begin{eqnarray}
\delta e'^{i}{_{\alpha}} &= \delta e^{i}{_{\alpha}} + \delta\mathcal{L}_{\vec{N}}e^{i}{_{\alpha}} = \delta e^{i}{_{\alpha}} + \mathcal{L}_{\vec{N}}\delta e^{i}{_{\alpha}},	\nonumber \\
\delta\omega'_{\alpha} &= \delta\omega_{\alpha} + \delta\mathcal{L}_{\vec{N}}\omega_{\alpha} = \delta\omega_{\alpha} + \mathcal{L}_{\vec{N}}\delta\omega_{\alpha},
\label{trans}
\end{eqnarray}
thus, $\bar{\omega}$ will under go the transformation
\begin{align}
\bar{\omega}' &= \int_{\Sigma}\epsilon^{\beta\alpha\gamma}\left(\delta\omega'_{\alpha}\wedge\delta\omega_{\gamma} + \Lambda\delta e'^{i}{_{\alpha}}\wedge e'_{i\gamma}\right)d\Sigma_{\beta} \nonumber	\\
&=\bar{\omega} + \int_{\Sigma}\mathcal{L}_{\vec{N}}\bar{\omega},
\end{align}
however, $\mathcal{L}_{\vec{N}}\bar{\omega} = \bar{\vec{N}}\cdot d\bar{\omega} + d(\vec{N}\cdot\bar{\omega})$, but $\delta\bar{\omega}=0$ (it is closed) and, the term $d(\vec{N}\cdot\bar{\omega})$ correspond to a surface term. Therefore $\bar{\omega}$ is invariant under infinitesimal diffeomorphisms.\\
In this manner, for reproducing the results previously stablished, we consider that upon picking $\Sigma$ to be the standard initial value surface $t=0$ the symplectic structure takes the standard  form
\begin{equation}
\bar{\omega} = \int_{\Sigma}\left(\delta\Pi^{a}\wedge\delta\omega_{a} + \delta P^{a}{_{i}}\wedge\delta e^{i}{_{a}}\right).	
\label{eq32a}
\end{equation}
where $\Pi^{a}\equiv\epsilon^{ab}\omega_{b}, P^{a}{_{i}}\equiv\Lambda\epsilon^{ab}e_{ib}$. In this manner, let $f$ any 0-form  defined on $Z$, hence the Hamiltonian vector field defined by the symplectic form (\ref{eq32a}) is given by
\begin{align}
X_{f}\equiv\int_{\Sigma}\frac{\delta f}{\delta\Pi^{a}}\frac{\delta}{\omega_{a}} - \frac{\delta f}{\delta\omega_{a}}\frac{\delta}{\delta\Pi^{a}} + \frac{\delta f}{\delta P^{a}{_{i}}}\frac{\delta}{\delta e^{i}{_{a}}} - \frac{\delta f}{\delta e^{i}{_{a}}}\frac{\delta}{\delta P^{a}{_{i}}}.
\label{vec}
\end{align}
Moreover, the Poisson bracket between two zero- forms is defined as usual
\begin{align}
\{f,g\}_P\equiv -X_f g &= \int_{\Sigma}\frac{\delta f}{\omega_{a}}\frac{\delta g}{\delta\Pi^{a}} - \frac{\delta f}{\delta\Pi^{a}} \frac{\delta g}{\delta\omega_{a}}+ \frac{\delta f}{\delta e^{i}{_{a}}}\frac{\delta g}{\delta P^{a}{_{i}}} - \frac{\delta f}{\delta P^{a}{_{i}}}\frac{\delta g}{\delta e^{i}{_{a}}}.
\end{align}
On the other hand, the definition of the vector field (\ref{vec}) requires that the constraints  (\ref{cons})  take a new fashion, this is,  they must be written in terms of the canonical momenta,   then   smearing the constraints  with test fields,  
\begin{align}
\gamma[N] &= \int_{\Sigma}N[\partial_{a}\Pi^{a} + \frac{1}{2\Lambda }\epsilon_{ij}\epsilon_{ab} P^{ai}P^{bj}],	\nonumber \\
\gamma_{i}[M^{i}] &= \int_{\Sigma}\frac{M^{i}}{\Lambda}[\partial_{a}P^{a}{_{i}} + \Lambda\epsilon_{ij} \epsilon_{ab}P^{aj}\Pi^{b}].
\label{cons2}
\end{align}
By inspection, the fundamental derivatives of the constraints (\ref{cons2}) are given by
\begin{eqnarray}
\frac{\delta\gamma}{\delta\omega_{a}} =0,	& \quad & \frac{\delta\gamma}{\delta\Pi^{a}} = -\partial_{a}N, 	\nonumber  \\
\frac{\delta\gamma}{\delta e^{i}{_{a}}} = \frac{N}{2}\epsilon_{ij}P^{aj},	&\quad &	 \frac{\delta\gamma}{\delta P^{a}{_{i}}} = \frac{N}{2}\epsilon^{i}{_{j}}e^{j}{_{a}},	\nonumber \\		
\frac{\delta\gamma_{i}}{\delta\omega_{a}} = 0,		&\quad & \frac{\delta\gamma_{i}}{\delta\Pi^{a}} = \Lambda M^{i}\epsilon_{ij}e^{j}{_{a}},                \nonumber 		\\
\frac{\delta\gamma_{i}}{\delta e^{i}{_{a}}} = \Lambda M^{j}\epsilon_{ji}\Pi^{a},	&\quad & \frac{\delta\gamma_{i}}{\delta P^{a}{_{i}}} = -\partial_{a}M^{i} + M^{i}\epsilon_{ij}\epsilon_{ca}\Pi^{a} = -\partial_{a}M^{i} - M^{i}\epsilon_{ij}\omega_{a},	\nonumber 	\\
\{\omega_{a}, \gamma\}_P = \partial_{a}N,	& \quad &  \{e^{i}_{a}, \gamma\}_P = N\epsilon_{j}{^{i}} e^j_a, 	\nonumber \\
\{\omega_{a}, \gamma_{i}\}_P = \epsilon_{ij}e^{j}{_{a}}M^{i}, & \quad &  \{e^{i}_{a}, \gamma_j\}_P = -\frac{\partial_aM^i}{\Lambda} + \frac{M^k}{\Lambda} \epsilon{^{i}}_k \omega_a, 
\label{36a}
\end{eqnarray}
In this manner, by taking into account (\ref{36a}), we observe that the motion generated by $\gamma[N]$ and $\gamma_{i}[M^{i}]$ is
\begin{align}
e'^{i}{_{a}}&\longrightarrow e^{i}{_{a}} + \varepsilon\partial_{a}M^{i} + \varepsilon M^{i}\varepsilon_{ij}\omega_{a} - \varepsilon N\Lambda \epsilon^{k}{_{j}}e^{j}{_{c}}+ O(\varepsilon^{2}),	\nonumber \\
\omega'_{a}&\longrightarrow\omega_{a} + \varepsilon\partial_{a}N + \varepsilon\epsilon_{ij}e^{j}{_{a}}M^{i} + O(\varepsilon^{2}).
\label{35}
\end{align}
Where $\epsilon$ is an infinitesimal parameter. Thus, we can observe that the gauge transformations $(\ref{35})$ are those found in previous sections. 
\section{Conclussions}
In this paper, the $HJ$ and the canonical covariant formalisms for 3dg without dynamics have  been performed. Within respect the $HJ$ method,  we have identified the complete set of involutive Hamiltonians, the characteristic equations,  the gauge transformations,  and  the counting of physical degrees of freedom was carried out. We reproduced the results obtained in \cite{8a} where   different  methods  were  used, in particular, we have showed that in the $HJ$ approach all dynamics of the system is given in terms of the fundamental differential;  in this respect  the $HJ$  method is more economic than either the Dirac or FJ methods.  On the other hand, we have completed our analysis by performing the canonical covariant method;  we have constructed a closed and gauge invariant symplectic structure from which we have identified the canonical structure of the theory and the gauge transformations. In this manner, with the present work, we have at hand a complete description for studying   gauge systems;  we can choose the Dirac,  FJ,  $HJ$ or the canonical covariant approaches, in any case we obtain generic results  and it  will be the base for studying gauge systems with a canonical structure more extensive than 3dg, for instance the Macdowell-€"Mansouri  formulation of gravity \cite{26, 27, 28},   all these ideas are in progress and will be reported soon \cite{29}.  

\end{document}